\documentclass[english,aps,preprint]{revtex4}
\usepackage{newtxtext,newtxmath}
\usepackage{mathrsfs}
\usepackage[export]{adjustbox}
\usepackage{physics}
\newtheorem{theorem}{Theorem}
\newtheorem{corollary}{Corollary}[theorem]
\newtheorem{lemma}{Lemma}

\usepackage{mathtools}
\DeclarePairedDelimiter\ceil{\lceil}{\rceil}

\usepackage{bbold}
\usepackage[colorlinks=true,citecolor=blue,breaklinks=true,linkcolor=red,
linktocpage=true,urlcolor=blue,pagebackref=false]{hyperref}

\usepackage{xcolor}

\begin{document}
		\title{Quantum activation functions for quantum neural networks}

	\author{Marco Maronese}
	\affiliation{Istituto di Fotonica e Nanotecnologie, Consiglio Nazionale delle Ricerche, Piazza Leonardo da Vinci 32, I-20133 Milano, Italy}
	\affiliation{Dipartimento di Fisica, Universit\'a degli Studi di Milano Bicocca, Piazza della Scienza, 3-20126 Milano, Italy}
	
	\author{Claudio Destri}
	\affiliation{Dipartimento di Fisica, Universit\'a degli Studi di Milano Bicocca, Piazza della Scienza, 3-20126 Milano, Italy}
	
	\author{Enrico Prati$^{*}$}
	\affiliation{Istituto di Fotonica e Nanotecnologie, Consiglio Nazionale delle Ricerche, Piazza Leonardo da Vinci 32, I-20133 Milano, Italy}
	\affiliation{*Corresponding Author: enrico.prati@cnr.it}
	
	

		\begin{abstract}
The field of artificial neural networks is expected to strongly benefit from recent developments of quantum computers.
In particular, quantum machine learning, a class of quantum algorithms which exploit qubits for creating trainable neural networks, will provide more power to solve problems such as pattern recognition, clustering and machine learning in general.
The building block of feed-forward neural networks consists of one layer of neurons connected to an output neuron that is activated according to an arbitrary activation function. The corresponding learning algorithm goes under the name of Rosenblatt perceptron. 
Quantum perceptrons with specific activation functions are known, but a general method to realize arbitrary activation functions on a quantum computer is still lacking.
Here we fill this gap with a quantum algorithm which is capable to approximate any analytic activation functions to any given order of its power series.
Unlike previous proposals providing irreversible measurement--based and simplified activation functions, here we show how to approximate any analytic function to any required accuracy without the need to measure the states encoding the information.   
Thanks to the generality of this construction, any feed-forward neural network may acquire the universal approximation properties according to Hornik's theorem.
Our results recast the science of artificial neural networks in the architecture of gate-model quantum computers. 
\end{abstract}
	
	\maketitle
	
	\maketitle
	\section{Introduction}
	

A quantum neural network encodes a neural network by the qubits of a quantum processor.  In
the conventional approach, biologically-inspired artificial neurons are implemented by
software as mathematical rate neurons.  For instance, the Rosemblatt perceptron (1957)
\cite{rosenblatt1957perceptron} is the simplest artificial neural network
consisting of an input layer of $N$ neurons and one output neuron behaving as a step
activation function. Multilayer perceptrons \cite{suter1990multilayer} are universal
function approximators, provided they are based on squashing functions. The latter consist of monotonic
functions which compress real values in a normalized interval, acting as activation
functions \cite{hornik1991approximation}.


In principle a quantum computer is suitable for performing tensor calculations typical of
neural network algorithms \cite{preskill2018quantum,aaronson2015read}.  Indeed,
the qubits can be arranged in circuits acting as layers of the quantum analogue of a
neural network.  If equipped with common activation functions such as the sigmoid and the
hyperbolic tangent, they should be able to process deep learning algorithms such those
used for problems of classification, clustering and decision making.  As qubits are
destroyed at the measurement event, in the sense that they are turned into classical bits,
implementing an activation function in a quantum neural network poses challenges requiring
a subtle approach. Indeed the natural aim is to preserve as much as possible the
information encoded in the qubits while taking advantage of each computation at the same
time. The goal therefore consists in delaying the measurement action until the end of the
computational flow, after having processed the information through neurons with a suitable
activation function.


Within the field of quantum machine learning
(QML)\cite{prati2017quantum,biamonte2017quantum}, if one neglects the implementation of quantum neural networks on adiabatic quantum computers \cite{rocutto2020quantum}, 
there are essentially two kind of
proposals of quantum neural networks on a gate-model quantum computer. The first consists
of defining a quantum neural network as a variational quantum circuit composed of
parameterized gates, where non-linearity is introduced by measurements operations
\cite{farhi2018classification,beer2020training,benedetti2019parameterized}. Such quantum neural networks
are empirically--evaluated heuristic models of QML not grounded on mathematical theorems
\cite{broughton2020tensorflow} Furthermore, this type of models based on variational quantum algorithms suffer from an exponentially vanishing gradient problem, the so-called barren plateau problem \cite{mcclean2018barren}, which requires some mitigation techniques \cite{grant2019initialization,cerezo2020cost}.  Quite differently, the second approach seeks to implement
a truly quantum algorithm for neural network computations and to really fulfill the
approximation requirements of Hornik's theorem \cite{hornik1989multilayer,hornik1991approximation} perhaps at the cost of a larger circuit depth.  Such
approach pertains to semi-classical \cite{daskin2018simple,torrontegui2019unitary}
or fully quantum \cite{cao2017quantum,hu2018towards} models whose non-linear
activation function is again computed via measurement operations.

Furthermore, quantum neural network proposals can be classified with respect to the
encoding method of input data.  Since a qubit consists of a superposition of the state $0$
and $1$, few encoding options are distinguishable by the relations between the number of
qubits and the maximum encoding capability.  The first is the $1$-\textit{to}-$1$ option
by which each and every input neuron of the network corresponds to one qubit
\cite{cao2017quantum,hu2018towards,da2016weightless,matsui2009qubit,da2016quantum}.  The most straightforward implementation
consists in storing the information as a string of bits assigned to classical base states
of the quantum state space.  A similar 1-{\it to}-1 method consists in storing a
superposition of binary data as a series of bit strings in a multi-qubit state.  Such
quantum neural networks are based on the concept of the quantum associative memory
\cite{ventura2000quantum,da2017neural}.  Another $1$-\textit{to}-$1$ option is
given by the quron (quantum neuron) \cite{schuld2014quest}.  A quron is a qubit whose $0$
and $1$ states stand for the resting and active neural firing state,
respectively\cite{schuld2014quest}.  

Alternatively, another encoding option consists in
storing the information as coefficients of a superposition of quantum states
\cite{shao2018quantum,tacchino2019artificial,kamruzzaman2019quantum,tacchino2020quantum,maronese2021continuous}.  The
encoding efficiency becomes exponential as an $n$-qubit state is an element of a
$2^n$-dimensional vector space.  To exemplify, the treatment by a quantum neural network
of a real image classification problem of few megabits makes the $1$-\textit{to}-$1$
option currently not viable \cite{pritt2017satellite}.  Instead, the choice $n$-\textit{to}-$2^n$ allows to encode a
megabit image in a state by using $\sim20$ qubits only.  

However, encoding the inputs as
coefficients of a superposition of quantum states requires an algorithm for generic
quantum state preparations \cite{shende2006synthesis,kuzmin2020variational,lazzarin2021multi} or, alternatively, to directly feed
quantum data to the network \cite{romero2017quantum}. For instance quantum encoding methods such as Flexible Representation of Quantum
Images (FRQI) \cite{le2011flexible} have been proposed.
Generally, to prepare an arbitrary $n$-qubit quantum state
requires a number of quantum gates that scales exponentially in $n$.  Nonetheless, in the
long run, an encoding of kind $n$-\textit{to}-$2^n$ guarantees a better applicability to
real problems than the options $1$-\textit{to}-$1$.  Moreover such encoding method
satisfies the requirements of Hornik's theorem in order to guarantee the universal
function approximation capability\cite{hornik1989multilayer}.  Despite some relatively
heavy constraints, such as the digital encoding and the fact that the activation function
involves irreversible measurements, examples towards this direction have been reported
\cite{tacchino2019artificial,tacchino2020quantum,maronese2021continuous}. Instead, differently from both the above proposals and from quantum annealing based algorithms applied to neural networks \cite{rocutto2020quantum}, we develop a fully reversible algorithm. 


In a novel alternative approach, we define here a $n$-to-$2^n$ encoding model that
involves inputs, weights and bias in the interval $\left[-1,1\right]\in \mathbb{R}$. The
model exploits the architecture of gate-model quantum computers to implement any
analytical activation function at arbitrary approximation only using reversible
operations.  The algorithm consists in iterating the computation of all the powers of the
inner product up to $d$-th order, where $d$ is a given overhead of qubits with respect to
the $n$ used for the encoding.  Consequently, the approximation of most common activation
functions can be computed by rebuilding its Taylor series truncated at the $d$-th order.


The algorithm is implemented in the \textit{QisKit }environment \cite{wille2019ibm} to build a one layer
perceptron with $4$ input neurons and different activation functions generated by power
expansion such as hyperbolic tangent, sigmoid, sine and swish function respectively,
truncated up to the 10-th order.  Already at the third order, which corresponds to the
least number of qubits required for a non-linear function, a satisfactory approximation of
the activation function is achieved.

This work is organized as follows: in Section 2, the definitions and the general strategy are summarized; in Section 3 the quantum circuits for the computation of the power terms and next of the the polynomial series are obtained. Next, in Section 4  the approximation of analytical activation functions algorithm is outlined while in Section 5 the computation of the amplitude is shown. 
Section 6 concerns the estimation of the perceptron output. The final Section is devoted to the conclusions.

\section{Definitions and general strategy}

	

	

\begin{figure}[t!]
		\centering
		\includegraphics[width=0.4\textwidth]{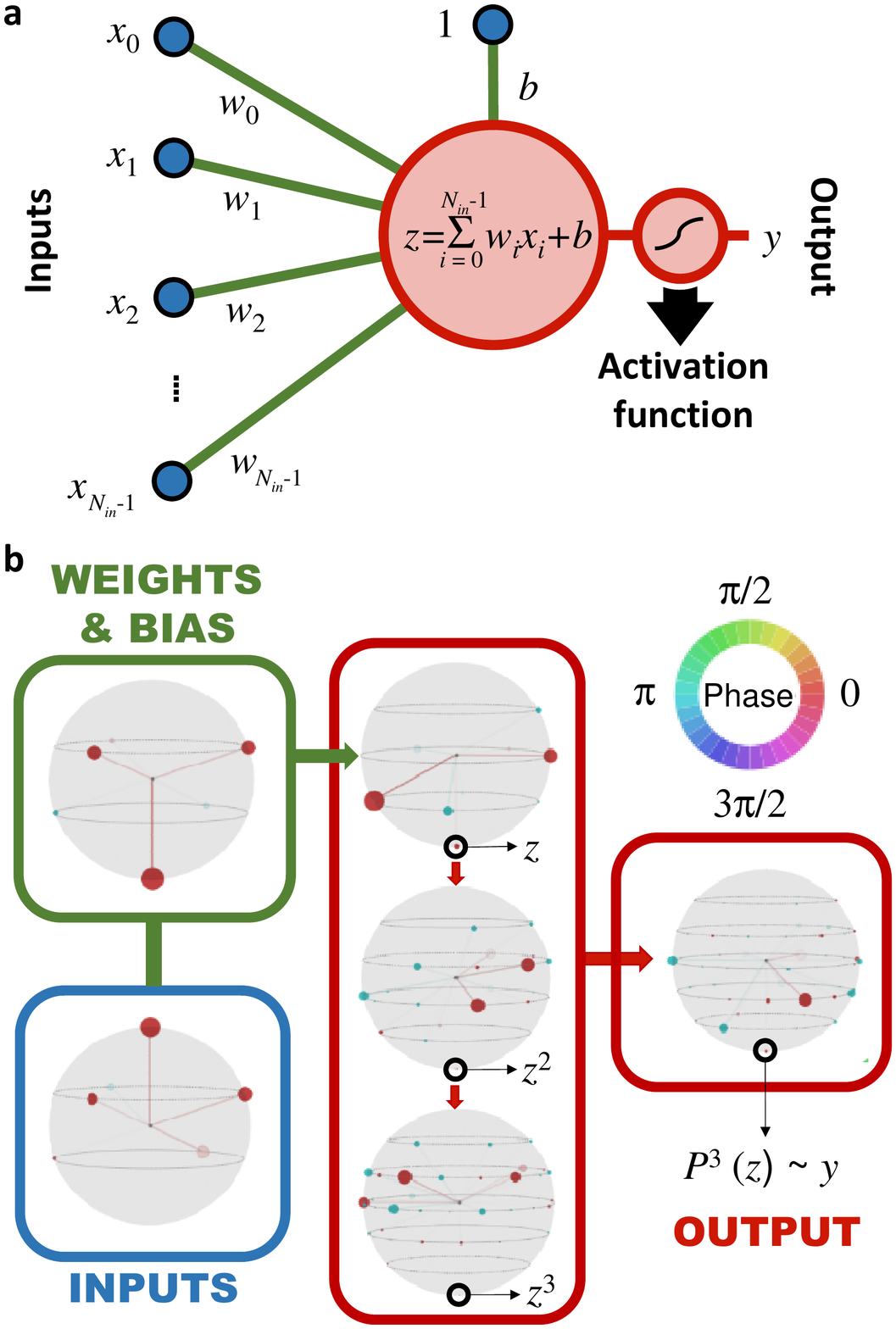}
		\caption{\textbf{Graphical representation of a one-layer perceptron and relative qubit-based version.} 
		A one-layer perceptron architecture (\textbf{a}) is composed by an input layer of $N_{in}$ neurons connected to the single output neuron. 
		It is characterized by an input vector $\vec x$, a weight's vector $\vec w$ and a bias $b$. 
		In the classical version, the activation function takes as argument the value $z=\vec w \cdot \vec x + b$ and it returns the perceptron output $y$. 
		The quantum version (\textbf{b}) follows the same architecture but the calculus consists of a sequence of transformations of a $n$-qubits quantum state initialized with the coefficients of the inputs vector $\vec x$ as probability amplitudes. 
		The quantum states at each step are represented by a qsphere (graphical representation of a multi-qubit quantum state). In a qsphere each point is a different coefficient of the superposition quantum state.
		Generally the coefficients are complex and in a qsphere the modules of the coefficients are proportional to the radius of the points while the phases depend on the colour. 
		In the blue box it is shown the starting quantum state with the inputs stored as probability amplitudes.
		Instead, in the green box it is shown the quantum state with the weights and the bias. 
		In the first red box, at each step one qubit is added in order to store the power terms of $z$ up to $d$ ($d=3$ in the Figure). 
		In the last red box the output of the perceptron is given by a series of rotations which compose a polynomial $P^d(z)$ ($d=3$ in the Figure).
		}
		\label{Fig1}
	\end{figure}

In order to define our quantum version of the perceptron with continuous parameters and arbitrary analytic activation function, let's consider a one-layer perceptron. The latter represents the fundamental unit of a feed-forward neural network. A one-layer perceptron is composed of $N_{in}$ input neurons and one output neuron equipped of an activation function $f:\mathbb{R}\to I$ where $I$ is a compact set.
The output neuron computes the inner product between the vector of the input values $\vec x = \left(x_0, x_1, \dots, x_{N_{in}-1}\right)\in \mathbb{R}^{N_{in}}$ and the vector of the weights $\vec w = \left(w_0, w_1, \dots, w_{N_{in}-1}\right)\in\mathbb{R}^{N_{in}}$ plus a bias value $b\in\mathbb{R}$.
Such scalar value is taken as the argument of an activation function. 
The real output value $y\in I$ of the perceptron is defined as  $y \equiv f\left(\vec w \cdot \vec x + b\right)$ as in Figure \ref{Fig1}\textbf{a}.

Here we develop a quantum circuit that computes an approximation of $y$. The algorithm starts by calculating the inner product $\vec w\cdot\vec x$ plus the bias value $b$. Next, it evaluates the output $y$ by calculating an approximation of the activation function $f$.
On a quantum computer, a measurement operation apparently represents the most straightforward implementation of a non-linear activation function, as done for instance in Ref. \cite{tacchino2019artificial} to solve a binary classification problem on a quantum perceptron. Such approach, however, cannot be generalized to build a multi-layered qubit-based feed-forward neural network.
First of all, measurement operations break the quantum algorithm and impose initialization of the qubits layer by layer, thus preventing a single quantum run of a multi-layer neural network. Secondly, other activation functions -- beside that implied by the measurement operations, are more suitable to solve generic problems of machine learning.

We avoid both of these shortcomings with a new quantum algorithm, which is based on two theorems as detailed below. 
The quantum algorithm is composed of two steps (Figure \ref{Fig1}\textbf{b}). First, the powers of $\vec w \cdot \vec x + b$ are stored as amplitudes of a multi-qubit quantum state. Next, the chosen activation function is approximated by building its polynomial series expansion through rotations of the quantum state. 
The rotation angles are determined by the coefficients of the polynomial series of the chosen activation function. They can be explicitly computed by our quantum algorithm.
Let's first summarize the notation used throughout the text.
Let $\mathcal{H}$ stand for the $2$-dimensional Hilbert space associated to one
qubit. Then the $2^n$-dimensional Hilbert space associated to a register $q$ of $n$ qubits
is written as $\mathcal{H}^{\otimes
  n}_q\equiv\mathcal{H}_{q_{n-1}}\otimes\mathcal{H}_{q_{n-2}}\otimes\ldots\otimes\mathcal{H}_{q_{0}}$. If
we denote by $\{\ket{0}, \ket{1}\}$ the computational basis in $\mathcal{H}$, then the
computational basis in $\mathcal{H}^{\otimes n}_q$ reads
$\{\ket{s_{n-1}s_{n-2}\ldots s_0},\;s_k\in\{0,1\}\hspace{1mm},\;
k=0,1,\ldots,n-1\}$.  An element $\ket{s_{n-1}s_{n-2}\ldots s_0}$ of this
computational basis can be alternatively written as $\ket{i}$ where
$i\in\{0,1,\ldots,2^n-1\}$ is the decimal integer number that corresponds to the bit
string $s_{n-1}s_{n-2}\ldots s_0$. In particular, if $N=2^n$, then
$\ket{N-1}\equiv \ket{2^n-1}\equiv\ket{11\ldots1}\equiv\ket{1}^{\otimes n}$.  In this
notation, the number of qubits of a register is indicated with a lowercase letter, such as
$n$ and $d$, while the dimension of the associated Hilbert space is indicated by the
correspondent uppercase letter, such as $N=2^n$ and $D=2^d$.

The expression $U_q^{\otimes
  n}=U_{q_{n-1}}\otimes U_{q_{n-2}}\otimes\cdots\otimes U_{q_{0}}$ represents a separable
unitary transformation constructed with one-qubit transformations $U_{q_j}$ acting on each
qubit of the register $q$.  A non-separable unitary multi--qubit transformation is usually
written as $U_q$ and, in some cases, simply $U$.
Two registers $a$ and $q$, respectively with $d$ and $n$ qubit, can be compound in a
single register supporting the $N+D$ Hilbert space $\mathcal{H}_a^{\otimes
  d}\otimes\mathcal{H}_q^{\otimes n}$  with computational basis
$\{\ket{i}_a\ket{j}_q,\,i=1,\ldots,D,\,j=1,\ldots,N\}$. For brevity, we will use the
compact notation $O_q$ for $\mathbb{1}_a\otimes O_q$ and $O_a$ for $\mathbb{1}_a\otimes
O_q$ for operators $O$ acting on only one of two registers. In particular, we write
\begin{equation*}
  \ketbra{i}_q  \equiv \mathbb{1}_a\otimes \ket{i}_q{_q}\!\bra{i}
\end{equation*}
for the $D-$dimensional projection onto the state $\ket{i}$ of the $q$ register. 

Particular cases of unitary operators implementable on a circuital model quantum computer are the controlled gates. Let $C_{q_i}U_{q_j}$ represent a controlled-$U$ transformation: the operator $U$ is applied on the qubit $q_j$ (called target qubit) if $q_i$ is in the state $\ket{1}$ (called control qubit). 
The transformation $\bar C_{q_i}U_{q_j}$ is a controlled transformation where the gate $U$ is applied on the qubit $q_j$ if $q_i$ is in the state $\ket{0}$. Therefore, $\bar C_{q_i}U_{q_j}=X_{q_i}C_{q_i}U_{q_j}X_{q_i}$.
In a more general case, a $d$-controlled operator has a notation of kind $C_{a}^{d}U_{q_j}$ where, in such case, $a$ is the set of the qubits control while $q_j$ is the target.


In the following, two qubit registers $q$ and $a$ of $n$ and $d$ qubits, respectively, are assumed to be assigned.


\section{Computation of the polynomial series}
As stated above, our aim is to build a $(n+d)$-qubits quantum state containing the Taylor expansion of $f(z)$ to order $d$, where $z= \vec w \cdot \vec x + b$ up to a normalization factor.
The number $n$ of required qubits, in addition to $d$, is determined by the dimension of the input vector.
We first need to encode the powers ($1,z,z^2,\dots,z^d$) in the $(n+d)$ qubits. The following Lemma provides the starting point:
\begin{lemma}
Given two vectors $\vec x, \vec w\in [-1,1]^{N_{in}}$ and a number $b\in[-1,1]$, and given a register of $n$ qubits such that $N=2^n\geq N_{in}+3$, then there exists a quantum circuit realizing a unitary transformation $U_z(\vec x, \vec w, b)$ such that
\begin{equation}
    \mel{N-1}{U_z(\vec x, \vec w, b)}{0} =\frac{\vec w \cdot \vec x +b}{N_{in}+1}\equiv z
    \label{eq:udot}
\end{equation}
where $\ket{0}\equiv\ket{0}^{\otimes n}$ and $\ket{N-1}\equiv\ket{1}^{\otimes n}$.
\label{lem:1}
\end{lemma}


In Lemma \ref{lem:1} a $n-$qubit unitary operator $U_z$ is defined by the requirement that
Eq. \ref{eq:udot} holds, where $b\in [-1,1]$, $\vec x=\left(x_0,\dots,x_{N_{in}-1}\right)$
and $\vec w=\left(w_0,\dots,w_{N_{in}-1}\right)$, where $ N_{in}\leq 2^n-3$ and $ x_i,
w_i\in [-1,1]$. The existence of infinitely many such operators is trivially obvious from
the purely mathematical point of view. The problem is to provide an explicit realization
in terms of
realistic quantum gates.
\\
\textbf{Proof:}
Let us define two vectors in $\mathbb{R}^{N}$: $\vec v_x=\left(\vec x, 1, A_x, 0 \right)$
and $\vec v_{w,b}=\left(\vec w, b, 0, A_{w,b} \right)$ where $ N\equiv 2^n$. In such
vectors $ N-N_{in}-3$ coefficients are always null while the values $ A_x$ and $ A_{w,b}$
are suitable constants defined such that 
$\vec v_x\cdot \vec v_x= \vec v_{w,b}\cdot\vec v_{w,b} = N_{in} + 1$.\\
It then follows that $\vec v_{w,b}^T\vec v_x=\vec w \vec x +b$ $\in$
$[-N_{in}-1, N_{in}+1]$.  We now define two $n$-qubit quantum states $\ket{\psi_x}$ and
$\ket{\psi_{w,b}}$ as follows
\begin{equation}
    \ket{\psi_x} =  \sum_{i=0}^{N-1} 
    \frac{v_{x,i}}{\sqrt{N_{in}+1}}\ket{i},\hspace{5mm}
    \ket{\psi_{w,b}} =
    \sum_{i=0}^{N-1}\frac{v_{w,b,i}}{\sqrt{N_{in}+1}}\ket{i} \;.
    \label{eq:newstate}
\end{equation}
Then, by construction 
\begin{equation*}
  \braket{\psi_x}{\psi_{w,b}} =  \frac{\vec w \cdot \vec x +b}{N_{in}+1}\equiv z \;.
\end{equation*}
The initialization algorithm mentioned above allows us to consider unitary transformations
$U_x=\mathcal{U}\left(\vec v_{x}\right)$ and $U_{w,b}=X^{\otimes
  n}\mathcal{U}^{\dag}\left(\vec v_{w,b}\right)$, where $X$ stands for the quantum NOT
gate, such that $U_x\ket{0}=\ket{\psi_x}$ and $ U_{w,b}\ket{\psi_{w,b}}=\ket{1}$. It follows that
\begin{equation}
\begin{split}
    \braket{\psi_{w,b}}{\psi_x}&=\mel{\psi_{w,b}}{U^{\dag}_{w,b}U_{w,b}}{\psi_x}=\mel{N-1}{U_{w,b}}{\psi_x}\\
    &=\mel{N-1}{U_{w,b}U_x}{0}^{\otimes n}
\end{split}
\end{equation}
Comparing with the equations \ref{eq:udot} we see that $U_z(\vec x, \vec w,
b)=U_{w,b}U_x=X^{\otimes n}\mathcal{U}^{\dag}\left(\vec
  v_{w,b}\right)\mathcal{U}\left(\vec v_{x}\right)$.
\\\qquad\hspace*{\fill}$\square$
\bigskip
\begin{figure}[t!]
  \centering
  \includegraphics[width=0.65\textwidth]{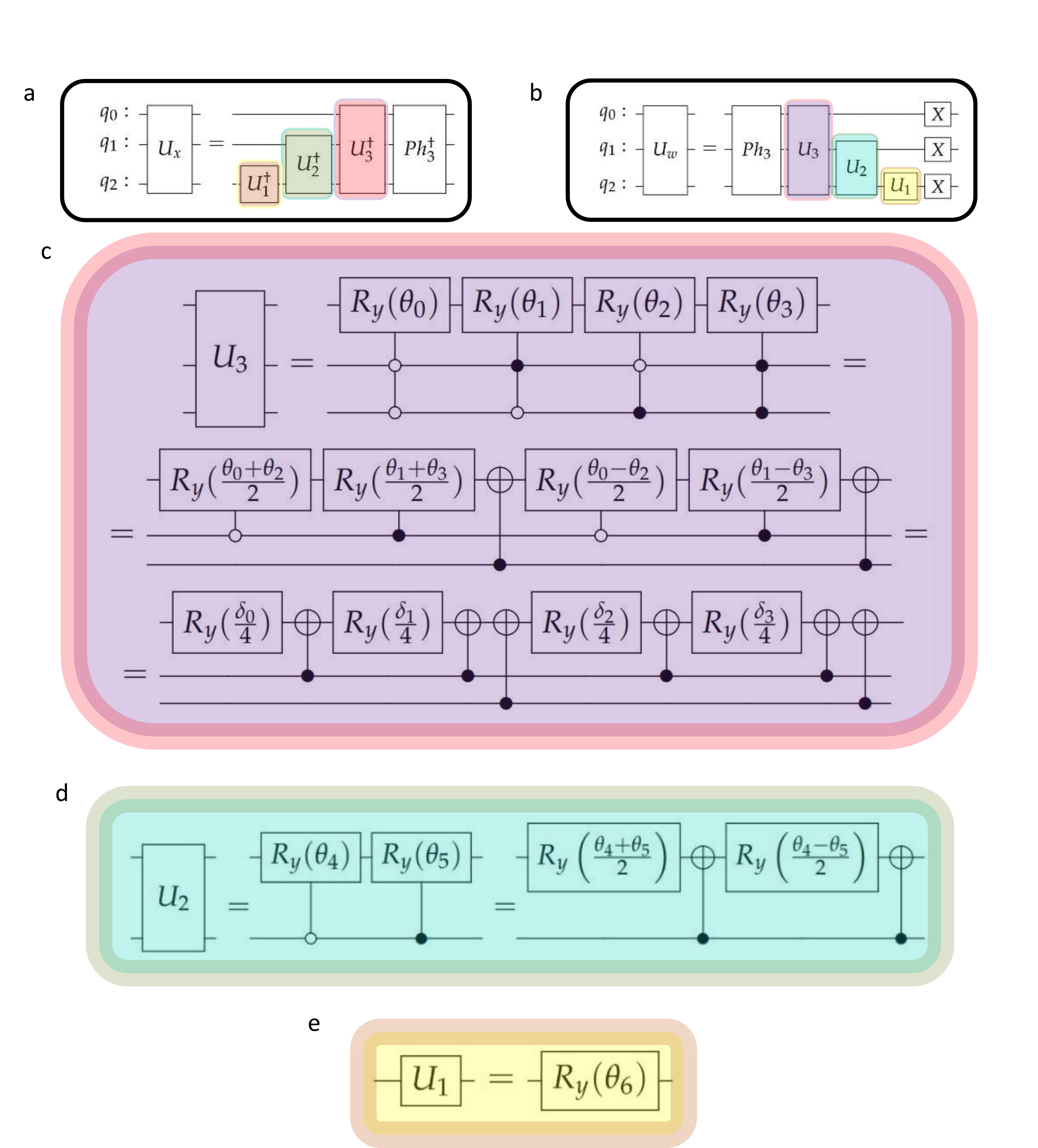}
  \caption{\textbf{Fundamental gates composition of the transformations $U_{w,b}$ and
      $U_{x}$}. The transformations $\mathcal{U}\left(\vec v_x\right)$ and
    $\mathcal{U}\left(\vec v_{w,b}\right)$ encode respectively the coefficients of the
    vectors $v_{x}$ and $v_{w,b}$ in a superposition quantum state. They are composed of
    the inverse of the operators $U_i$ where $i=1,\dots,n$ and $Ph_3$, which introduces
    the phases of the probability amplitudes. The operators $U_3$ \textbf{c)}, $U_2$
    \textbf{d)} and $U_1$ \textbf{e)} are shown as composition of multi-controlled
    rotations $R_y$ which are equal to a composition of gates $R_y$ and Controlled-Not
    \cite{mottonen2005decompositions}. The transformation $\scriptstyle Ph^{\dag}_3$
    (\textbf{a}) introduces the phases of the amplitudes while $Ph_3$(\textbf{b})
    removes them. The details of the arbitrary quantum state preparation circuit are
    described in Supplementary Note \textbf{1}.}
		\label{Fig3}
	\end{figure}

Since the amplitudes of the states $\ket{\psi_x}$ and $\ket{\psi_{w,b}}$ are real, the phases are either $0$ or $\pi$ and it is no longer necessary to apply a series of multi-controlled $R_Z$ to set them.
A single diagonal transformation suffices, with either $1$ or $-1$ on the diagonal.
For such purpose, hypergraph states prove effective \cite{rossi2013quantum}. Thanks to such kind of states, a small number of Z, CZ, and multi-controlled Z gates are needed to achieve the transformation $\mathcal{U}\left(\vec v\right)$. The transformations, which introduce the phases of the amplitudes of a $n$-qubits quantum state, are summarized by an operator called  $Ph^{\dag}_n$ in the Figure \ref{Fig3}. 
More details about the strategy adopted for quantum--state initialization is reported in Supplementary Note \textbf{1} and Supplementary Note \textbf{2}.\\
There are many alternatives to the states $\ket{\psi_x}$ and $\ket{\psi_{w,b}}$ which give the same inner product $\braket{\psi_{w,b}}{\psi_x}=z$.
Defining the two vectors
\begin{equation}
\begin{split}
    &\vec v_x = \left(A_x, x_0, \dots, x_{N_{in}-1},1,0,\dots,0,0\right)\in \mathbb{R}^{N}\\
    &\vec v_{w,b} = \left(0, w_0, \dots, w_{N_{in}-1},b,0,\dots,0,A_{w,b} \right)\in \mathbb{R}^{N}
\end{split}    
\end{equation}
then the transformations $\mathcal{U}\left(\vec v_x\right)$ and $\mathcal{U}\left(\vec v_{w,b}\right)$ applied on the state $\ket{0}^{\otimes n}$ return two states, $\ket{\psi_x}$ and $\ket{\psi_{w,b}}$ respectively, such that $\braket{\psi_{w,b}}{\psi_x}=z$.
The reason for the choice shown above is due to the phases to add. 
Since the values $A_{w,b}$ and $A_x$ do not appear in the inner product then their phases are not relevant. 
Therefore, such states $\ket{\psi_x}$ and $\ket{\psi_{w,b}}$, make unnecessary a $(n-1)$-controlled Z gate to adjust the phases of the amplitudes associated with $\ket{0}$ and $\ket{N-1}$.
In the Figure \ref{Fig3} the composition of the transformations $U_x$ (Fig.\ref{Fig3}\textbf{-a}) and $U_{w,b}$ (Fig. \ref{Fig3}\textbf{-b}) are shown in the case of a one-layer perceptron with $N_{in}=4$ neurons. In such case the number of input neurons is $4$. Since $n = log_2N$ and $N\geq N_{in}+3$ then, given $N_{in}$ input neurons the minimum number of required qubits is $n=\ceil{\log_2(N_{in}+3)}$.
Therefore, with $N_{in}=4$,  n=3 qubits are required to store $z$ in a quantum state. 


The variable $z$ generalizes in two respects the inner product of  Ref. \cite{tacchino2019artificial} where
inputs and weights only take binary values $\{-1,1\}$ and no bias is involved.

\begin{figure*}[t!]
  \centering
  \includegraphics[scale=.3]{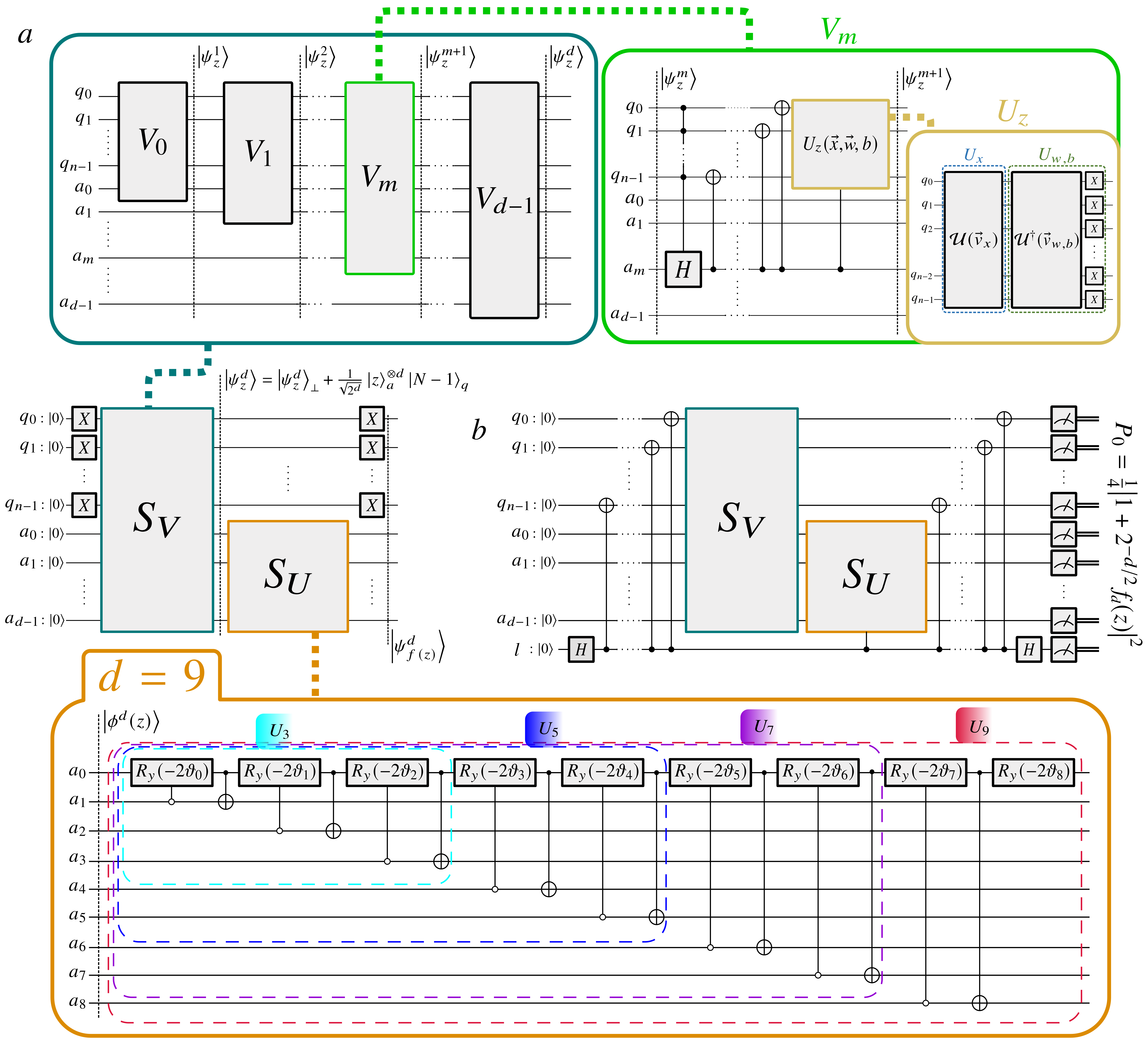}
		\caption{\textbf{Quantum circuit of a qubit-based one-layer perceptron in two different cases.} 
		\textbf{a)} The quantum circuit which returns the state $|\psi^d_{f(z)}\rangle$ which encodes the value $f_d(z)$ as stated by the Theorems \ref{th:3teo} and \ref{th:teof}. 
		In a more general context $f_d(z)$ is the output value of a neuron of a deep neural network hidden layer, as no measurement is required when the information is sent to the next layer. 
		The subroutines $S_V$ and $S_U$ are described in the upper boxes, where the general case of $S_V$ is shown, and in the lower box respectively. In the latter, it is shown a particular case at $d=9$. Such value corresponds to the maximum order of $d=9$ of the polynomial expansion used for approximating the activation functions \textit{tanh}, \textit{sigmoid} and \textit{sine}.
		\textbf{b)} A complete quantum circuit of a qubit-based one-layer perceptron is shown. In such case, by performing the measurements of all the qubits and averaging after many repetitions of the circuit, it is possible to estimate the output $y_q$ of the one-layer perceptron. 
		}
	\label{ref:fig4}
\end{figure*}

The transformation $U_z(\vec x, \vec w, b)$ is a key building block of the quantum perceptron algorithm. Indeed, in our quantum circuit such transformation is iterated several times over the Hilbert space enlarged to $\mathcal{H}^{\otimes d}_a\otimes \mathcal{H}_q^{\otimes n}$ by the addition of another register $a$ of $d$ qubits.
The existence of such a quantum circuit is guaranteed by the following theorem that provides its explicit construction:
\begin{theorem}\label{th:3teo}
Let $z$ be the real value in the interval $\left[-1,1\right]$ assumed by $\left(\vec w \cdot \vec x + b\right)/\left(N_{in}+1\right)$, where $\vec x, \vec w\in [-1,1]^{N_{in}}$ and $b\in[-1,1]$. 
Let $q$ and $a$ be two registers of $n$ and $d$ qubits respectively, with $N=2^n\geq N_{in}+3$.
Then there exists a quantum circuit which transforms the two registers from the initial state  $\ket{0}_a\ket{0}_q$ to a $(n+d)$-qubit entangled state $|\psi^d_{z}\rangle$ 
of the form
\begin{equation}\label{eq:defpsid}
 |\psi^d_{z}\rangle = |\psi_z^d\rangle_{\perp} + 
    \frac{1}{2^{d/2}}\ket{z}^{\otimes d}_a\,\ket{N-1}_q \;,
\end{equation}
where
\begin{equation*} 
    \ketbra{N-1}_q|\psi_z^d\rangle_{\perp} = 0 
\end{equation*}
and
\begin{equation*} 
  \ket{z}\equiv\ket{0}+z\ket{1}\;.
\end{equation*}
The circuit is expressed by $S_VX^{\otimes n}_q$ (Fig. \ref{ref:fig4}\textbf{a}) where $X$
is the quantum NOT gate and 
\begin{equation*}
    S_V=V_{d-1}\cdots V_{1}V_{0}
\end{equation*}
with
\begin{equation*}\label{eq:master}
    V_m=C_{a_m}U_z(\vec x, \vec w, b)_q C_{a_m}X^{\otimes n}_q C_q^n H_{a_m}
    \;,\quad  m=0,1,\dots,d-1 \;. 
\end{equation*}
\end{theorem}
\textbf{Proof:} The thesis of the theorem is the existence of a transformation which, acting on two registers of qubit $q$ and $a$ with $n$ and $d$ qubits respectively, returns a state $\ket{\psi^d_z}\in \mathcal{H}_a^{\otimes d}\otimes \mathcal{H}_q^{\otimes n}$ as defined in the Equation \ref{eq:defpsid}. The demonstration consists of the construction of such a circuit.
For such purpose, let's define the $d$ states
$\ket{\psi_z^m}\in \mathcal{H}_{a_{m-1}}\otimes\mathcal{H}_{a_{m-2}}\otimes\cdots\otimes\mathcal{H}_{a_{0}}\otimes \mathcal{H}_q^{\otimes n}$, where $m=0,\dots,d-1$
\begin{equation}\label{eq:defpsi}
\begin{split}
     |\psi^m_{z}\rangle =& |\psi_z^m\rangle_{\perp}+|\psi_z^m\rangle_{\parallel} =\\
     =&|\psi_z^m\rangle_{\perp} + 
    \frac{1}{2^{m/2}}\ket{z}^{\otimes m}\,\ket{N-1}_q \;,
\end{split}
\end{equation}
where $\ketbra{N-1}_q|\psi_z^m\rangle_{\perp}=0\;\forall\, m$.\\
From such definition it follows that the states $\ket{\psi_z^m}$ are states of $(n+m)$-qubits and $\ket{\psi_z^0}\equiv \ket{N-1}_q$ is a $n$-qubits state.\\
The proof of the theorem is therefore reduced to demonstrating the existence of a sequence of transformations $V_m$, where $m=0,\dots,d-1$, such that $ V_m\ket{0}_{a_m}\ket{\psi_z^m}=\ket{\psi_z^{m+1}}$ where $a_m$ is the $m$-th qubit in the register $a$. 
Therefore $V_m$ is a unitary transformation defined over the space $\mathcal{H}_{a_m}\otimes\mathcal{H}_{a_{m-1}}\otimes\cdots\otimes\mathcal{H}_{a_{0}}\otimes \mathcal{H}_q^{\otimes n}$.\\
Let's consider the following ansatz for the transformation $V_m$:
\begin{equation}
    V_m=C_{a_m}U_z(\vec x, \vec w, b)_q C_{a_m}X^{\otimes n}_q C_q^n H_{a_m}
    \label{eq:master2}
\end{equation}
whose graphical representation is given in Figure \ref{ref:fig4}\textbf{a}.

Let's apply $ V_m$, as defined in the Equation \ref{eq:master2}, on the state $\ket{0}_{a_m}\ket{\psi_z^m}$. 
\begin{equation}
    \begin{split}
        &C_{a_m}U_z(\vec x, \vec w, b)_q C_{a_m}X^{\otimes n}_q C_q^n H_{a_m}\ket{0}_{a_m}\ket{\psi_z^m}=\\
        =&C_{a_m}U_z(\vec x, \vec w, b)_q C_{a_m}X^{\otimes n}_q\Bigl[\ket{0}_{a_m}\ket{\psi_z^m}_{\perp}+\frac{1}{\sqrt{2}}\left(\ket{0}_{a_m}+\ket{1}_{a_m}\right)\ket{\psi_z^m}_{\parallel}\Bigr]
    \end{split}
\end{equation}
The transformation $ C_{a_m}U_z(\vec x, \vec w, b)_q C_{a_m}X^{\otimes n}_q$ consists in the application of $ U_z(\vec x, \vec w, b)X^{\otimes n}$ on the qubits $q$ controlled by the qubit $a_m$ which means the transformation act only on $\ket{\psi_z^m}_{\parallel}$ so focussing only on its subspace it results 
\begin{equation}
        U_z(\vec x, \vec w, b)_q X^{\otimes n}_q\ket{\psi_z^m}_{\parallel}
        =\frac{1}{2^{m/2}}\ket{z}^{\otimes m}U_z(\vec x, \vec w, b)\ket{0}^{\otimes n}_q
\end{equation}
Therefore the transformation $ V_m$ applied on $\ket{0}_{a_m}\ket{\psi_z^m}$ returns the following state
\begin{equation}
    \begin{split}
        &V_m\ket{0}_{a_m}\ket{\psi_z^m}=\ket{0}_{a_m}\ket{\psi_z^m}_{\perp}+\\
        &+\frac{1}{\sqrt{2}}\biggl(\ket{0}_{a_m}\frac{1}{2^{m/2}}\ket{z}^{\otimes m}\ket{N-1}_q+\ket{1}_{a_m}\frac{1}{2^{m/2}}\ket{z}^{\otimes m}U_z(\vec x, \vec w, b)\ket{0}^{\otimes n}_q\biggr)
    \end{split}
    \label{eq:statoth}
\end{equation}
To demonstrate that the state just obtained is $\ket{\psi_z^{m+1}}$, the projection over the state $\ket{N-1}_q$ must return $\frac{1}{\sqrt{2^{m+1}}}\ket{z}^{\otimes (m+1)}\ket{N-1}_q$ as from the definition of the states $\ket{\psi_z^{m}}$. Let's apply the projection $ \ketbra{N-1}{N-1}_q$ on the resulting state in the Equation \ref{eq:statoth}.
Since $\ketbra{N-1}{N-1}_q\ket{\psi_z^m}_{\perp}=0$ by definition, and $\mel{N-1}{U_z(\vec x, \vec w, b)}{0}^{\otimes n}_q=z$ as from Lemma \ref{lem:1}, the result of the projection $ \ketbra{N-1}{N-1}_q$ is the following.
\begin{equation}
    \begin{split}
        &\frac{1}{\sqrt{2}}\biggl(\ket{0}_{a_m}\frac{1}{\sqrt{2^m}}\ket{z}^{\otimes m}\ket{N-1}_q+z\ket{1}_{a_m}\frac{1}{2^{m/2}}\ket{z}^{\otimes m}\ket{N-1}_q\biggr)=\\
        =&\frac{1}{2^{m+1}}\ket{z}^{\otimes (m+1)}\ket{N-1}_q=\ket{\psi_z^{m+1}}_{\parallel}
    \end{split}
\end{equation}
Having demonstrated that $ V_m\ket{0}_{a_m}\ket{\psi_z^m}=\ket{\psi_z^{m+1}}$, the proof of the existence of the transformation which returns $\ket{\psi_z^d}$ if applied on $\ket{0}_a^{\otimes d}\ket{0}_q^{\otimes n}$ proceeds by recursion. Indeed, by applying $ V_{d-1}\cdots V_1 V_0$ to the state $\ket{0}_a^{\otimes d}\ket{\psi_z^0}=\ket{0}_a^{\otimes d}\ket{N-1}_q$ the resulting state will be $\ket{\psi_z^d}$.\\\qquad\hspace*{\fill}$\square$

To summarize,
the quantum circuit of the quantum perceptron algorithm starts by expressing the unitary
operator which initializes the $q$ and $a$ registers from the state $\ket{0}_a\ket{0}_q$ to the
state $\ket{\psi^d_{z}}$.  Such a unitary operator is expressed by $S_V X^{\otimes n}$
where $S_V$ is the subroutine of the quantum circuit which achieves the goal of the first
step of the quantum perceptron algorithm, i.e. to encode the powers of $z$ up to $d$ in a
quantum state as from the following Corollary. 
\begin{corollary}\label{co:co1}
The state $|\psi_z^d\rangle$ stores as probability amplitudes all the powers $z^k$, for $k=0,1,\ldots,d$, up to a trivial factor. Indeed Eq.~\ref{eq:defpsid} in Theorem \ref{th:3teo} implies
\begin{equation}
    _q\!\bra{N-1}\,_a\!\braket{2^k-1}{\psi_z^d}=
    2^{-d/2}z^k \;,\quad k=0,1,\ldots,d \;.
\end{equation}
\end{corollary}

The first step of the quantum perceptron algorithm consists of the storage of all powers of $z \equiv \left(\vec w \cdot \vec x + b\right)/\left(N_{in}+1\right)$ up to $d$ in a $(n+d)$-qubits state.
The proof of the Theorem \ref{th:3teo} implies that the first step of the algorithm is the quantum circuit shown in the Figure \ref{ref:fig4}\textbf{-a}, consisting of a subroutine composed by a Pauli gate $X$ applied on each qubit in the register $q$ and a transformation $S_V=V_{d-1}\cdots V_0$. Indeed from the Corollary \ref{co:co1} the state $\ket{\psi_z^d}=S_V X_q^{\otimes n}\ket{0}_a^{\otimes d}\ket{0}_q^{\otimes n}$ stores as probability amplitudes all the powers of $z$ up to $d$ less than a factor $2^{-d/2}$. 
The proof of the Corollary \ref{co:co1} is straightforward as follows.\\
\textbf{Proof:} As shown above, the state $\ket{\psi_z^d}$ can be written as $\ket{\psi_z^d}=\ket{\psi_z^d}_{\perp}+\ket{\psi_z^d}_{\parallel}$ where $\ketbra{N-1}{N-1}_q\ket{\psi_z^d}_{\perp}=0$, therefore, $_q\bra{N-1}_a\braket{2^k-1}{\psi_z^d}= _q\bra{N-1}_a\braket{2^k-1}{\psi_z^d}_{\parallel}$.\\ Since
$\ket{\psi_z^d}_{\parallel}=\frac{1}{\sqrt{2^d}}\ket{z}_a^{\otimes d}\ket{N-1}_q$ then
\begin{equation}
\begin{split}
    _q&\bra{N-1}_a\braket{2^k-1}{\psi_z^d}_{\parallel}=\\
    =&\frac{1}{\sqrt{2^d}}\braket{2^k-1}{z}^{\otimes d}_a\braket{N-1}{N-1}_q
\end{split}
\end{equation}
Let's rewrite $\ket{2^k-1}$ in a binary form $\ket{s_{d-1}s_{d-2}\cdots s_0}$ where $s_{j}=1$ from $j=0$ to $j=k-1$ and $0$ otherwise. 
\begin{equation}
    \frac{1}{\sqrt{2^d}}\braket{s_0,s_1,\dots,s_{d-1}}{z}_a^{\otimes d}=\frac{1}{\sqrt{2^d}}\prod_{i=1}^d\braket{s_{d-i}}{z}_{a_{d-i}}=2^{-d/2}z^k
\end{equation}
The latter holds because, $\forall j=0,\dots,d-1$,
\begin{equation}
\braket{s_{j}}{z}_{a_{j}}=\braket{s_{j}}{0}_{a_{j}}+z\braket{s_{j}}{1}_{a_{j}}=z^{s_{a_j}}
\end{equation}
Therefore, $_q\bra{N-1}_a\braket{2^k-1}{\psi_z^d}=2^{-d/2}z^k$.
\\\qquad\hspace*{\fill}$\square$

The next step of the algorithm consists in transforming the state $|\psi_z^d\rangle$ so as to achieve a special recursively defined $d$-degree polynomial in $z$.
Such step is identifiable with the subroutine $S_U$ of the quantum perceptron circuit, see Figure \ref{ref:fig4}\textbf{a}.\\
By Eq.~\ref{eq:defpsid} in Theorem \ref{th:3teo} there must 
exists a unitary operator $S_U$ which acts as the identity on $\mathcal{H}_q$ and returns, when applied to $\ket{z}^{\otimes d}_a$, a new state which stores the polynomial. In fact, it holds  the following



\begin{theorem}
Let $\{f_k,\, k = 1,\ldots,d\}$ be the family of polynomials in $z$ defined by the following recursive law
\begin{equation}
    f_k(z)=f_{k-1}(z)\cos{\vartheta_{k-1}}-z^k\sin{\vartheta_{k-1}} \;,\quad k = 1,\ldots,d \;,
\label{eq:ric2}
\end{equation}
with $f_0(z) = 1$ and $\vartheta_k\in
\left[-\frac{\pi}{2},\frac{\pi}{2}\right]$ for any $k = 0,\ldots,d-1$.\\
Then there exists a family $\{U_k,\,k = 1,\ldots,d\}$ of unitary operators such that
\begin{equation}
    _a\!\mel{0}{U_k}{z}^{\otimes d}_a=f_k(z) \;.
\end{equation}
These unitary operators are, in turn, defined by the recursive law
\begin{equation}\label{eq:Ufase2}
    U_k=C_{a_0}X_{a_k}\Bar{C}_{a_k}R_y(-2\vartheta_{k-1})_{a_0}U_{k-1} \;,\quad  k = 1,\ldots,d \;,
\end{equation}
with $U_0=\mathbb{1}$.
\label{th:teof}
\end{theorem}
The subroutine $S_U$ shown in Figure \ref{ref:fig4}\textbf{a} corresponds to $U_d$. The Proof of Theorem \ref{th:teof}
follows.
\textbf{Proof:} The proof of the theorem follows two steps, namely the statement for the first term of the polynomial and an inductive step  as follows. 
The first step consists of demonstrating that 
\begin{equation*}
_a\mel{0}{U_1}{z}_a^{\otimes d}=\cos{\theta_{0}}-z\sin{\theta_{0}}=f_1(z)
\end{equation*}
with $U_1=C_{a_0}X_{a_1}\Bar{C}_{a_1}R_y(-2\vartheta_{0})_{a_0}$ as defined in the Equation \ref{eq:Ufase2}.
In the second step, instead, the proof proceeds for $U_k\hspace{2mm}\forall k=1,2,\dots,d$ recursively.\\
It aims to prove $_a\mel{0}{U_k}{z}_a^{\otimes d}=f_{k}(z)$
assuming that $_a\mel{0}{U_{k-1}}{z}_a^{\otimes d}=f_{k-1}(z)$, where
\begin{equation*}
    U_k=C_{a_0}X_{a_k}\Bar{C}_{a_k}R_y(-2\vartheta_{k-1})_{a_0}U_{k-1})
\end{equation*}
as defined in the Equation \ref{eq:Ufase2}.
Let's preliminary consider the states $U_k\ket{z}^{\otimes d}_a$, where $d\geq k\geq 1$. The state $\ket{z}^{\otimes d}_a$ is considered in the case with $k=0$. Next, let's focus on the subspace of $\mathcal{H}^{\otimes d}_a$ defined as $\mathcal{H}_{\{0,1\}}=\{\ket{0}^{\otimes d},\ket{0}^{\otimes (d-1)}\ket{1}\}$. 
The operator which projects the elements of $\mathcal{H}^{\otimes d}_a$ in the subspace $\mathcal{H}_{\{0,1\}}$ is $\mathcal{P}_{\{0,1\}}=\ketbra{0}{0}_a+\ketbra{1}{1}_a$.\\
Let's now move to the first step of the demonstration. The first operation consists of applying $U_1$ to the state $\ket{z}^{\otimes d}_a$.
Because of the definition of the state $\ket{z}^{\otimes d}_a$, it follows that $\braket{2^i-1}{z}^{\otimes d}_a=z^i$ where $i=0,\dots,d$ (Corollary \ref{co:co1}), therefore, the projection over $\mathcal{H}_{\{0,1\}}$ of the state $\ket{z}^{\otimes d}_a$ is 
\begin{equation*}
    \mathcal{P}_{\{0,1\}}\ket{z}^{\otimes d}_a=\ket{0}^{\otimes d}+z\ket{0}^{\otimes (d-1)}\ket{1}
\end{equation*}
The operator $\Bar{C}_{a_1}R_y(-2\vartheta_0)_{a_0}=X_{a_1}C_{a_1}R_y(-2\vartheta_0)_{a_0}X_{a_1}$ rotates the qubit $a_0$, along the $y$-axis of the Bloch sphere of angle $-2\vartheta_0$, only if the qubit $a_1$ is in $\ket{0}$, therefore, such operator acts on the subspace $\mathcal{H}_{\{0,1\}}$.\\ 
The projection on such subspace of the state $\Bar{C}_{a_1}R_y(-2\vartheta_0)_{a_0}\ket{z}^{\otimes d}_a$ is
\begin{equation*}
    \begin{split}
        &(\cos{\theta_{0}}-z\sin{\theta_{0}})\ket{0}^{\otimes d}+\\
        +&(\sin{\theta_{0}}+z\cos{\theta_{0}})\ket{0}^{\otimes (d-1)}\ket{1}=\\
        =&f_1(z)\ket{0}^{\otimes d}+(\sin{\theta_{0}}+z\cos{\theta_{0}})\ket{0}^{\otimes (d-1)}\ket{1}
    \end{split}
\end{equation*}
Since $C_{a_0}X_{a_1}$ is a controlled-NOT gate which acts only if the qubits $a_0$ is in the state $\ket{1}$ then
\begin{equation*}
    \begin{split}
       &{_a}\mel{0}{U_1}{z}^{\otimes d}_a={_a}\mel{0}{C_{a_0}X_{a_1}\Bar{C}_{a_1}R_y(-2\vartheta_{0})_{a_0}}{z}=\\
       =&\cos{\theta_{0}}-z\sin{\theta_{0}}=f_1(z) 
    \end{split}
\end{equation*}
which completes the first step of this demonstration.
Let's now demonstrate the recursive step. Here, the only assumption is $_a\mel{0}{U_{k-1}}{z}^{\otimes d}_a=f_{k-1}(z)$, therefore, differently from the previous step where the projection of $\ket{z}^{\otimes d}_a$ on the subspace $\mathcal{H}_{\{0,1\}}$ was known, here the projection of $U_{k-1}\ket{z}^{\otimes d}_a$ is equal to 
\begin{equation}
    f_{k-1}(z)\ket{0}^{\otimes d}+B_{k-1}\ket{0}^{\otimes (d-1)}\ket{1}
    \label{eq:as}
\end{equation}
where $B_{k-1}$ is an unknown real value.
Let's apply $C_{a_0}X_{a_k}\Bar{C}_{a_k}R_y(-2\vartheta_{k-1})_{a_0}$ on the state $U_{k-1}\ket{z}^{\otimes d}_a$ so as to obtain the state $U_{k}\ket{z}^{\otimes d}_a$.
From the Equation \ref{eq:as}:
\begin{equation*}
    _a\mel{0}{U_{k}}{z}^{\otimes d}_a=f_{k-1}(z)\cos{\vartheta_{k-1}}-B_{k-1}\sin{\vartheta_{k-1}}
\end{equation*}
To prove the theorem, $B_{k-1}$ must be equal to $z^k$ since $f_{k}(z)=f_{k-1}(z)\cos{\vartheta_{k-1}}-z^k\sin{\vartheta_{k-1}}$.\\
The purpose of the second step of the proof can be achieved just proving that $B_{k-1}=z^k$ $\forall k=1,\dots,d$. 
That is already proved for $k=0$ because $\braket{2^i-1}{z}^{\otimes d}_a=z^i$ as said above.
Let's prove that $B_{k-1}=z^k$ for $k=1$ while for $k>1$ the proof will proceed recursively.\\
The state $\ket{2^i-1}$ is a state of the computational bases of $\mathcal{H}^{\otimes d}_a$. Writing such state in the binary version it results equal to $\ket{s_{d-1}s_{d-2}\cdots s_0}$ where $s_{j}=1$ from $j=0$ to $j=i-1$ and $0$ otherwise.\\
As said before the operator
$\Bar{C}_{a_1}R_y(-2\vartheta_0)_{a_0}$ acts only on the state $\ket{s_{d-1}s_{d-2}\cdots s_0}$ where $s_{1}=0$, therefore, it does not act on the states $\ket{2^i-1}$ $\forall i>1$. 
Instead, the operator $C_{a_0}X_{a_1}$ acts on the state $\ket{s_{d-1}s_{d-2}\cdots s_0}$ where $s_{0}=1$ and it applies a NOT operation on the bit $s_1$. 
That means the states $\ket{2^i-1}$ become $\ket{2^i-1-2}$ $\forall i>1$. Therefore, since $\braket{2^i-1}{z}=z^i$, thanks to $U_{1}(\vartheta_{0})=C_{a_0}X_{a_1}\Bar{C}_{a_1}R_y(-2\vartheta_{0})_{a_0}$ then $\mel{2^i-1-2}{U_{1}}{z}^{\otimes d}_a=z^i$ $\forall i> 1$. 
In particular, taking the value of $i$ such that $\ket{2^i-1-2}=\ket{0\cdots01}$, that is $i=2$, $\mel{0\cdots01}{U_{1}}{z}^{\otimes d}_a=z^2$ and, therefore, $B_1=z^2$.
Let's proceed recursively for $k>1$ assuming that 
\begin{equation}
    \mel{\scriptstyle{2^i-1-\sum_{h=1}^{k-1}2^{h}}}{U_{k-1}(\vec\vartheta_{k-2})}{z}=z^i \hspace{2mm}\text{where} \hspace{2mm}d>i\geq k
    \label{eq:as2}
\end{equation}
The state $\ket{\scriptstyle{2^i-1-\sum_{h=1}^{k-1}2^{h}}}$ written in a binary form is $\ket{s_{d-1}s_{d-2}\cdots s_0}$ where $s_{j}=1$ from $j=k$ to $j=i-1$ and for $j=0$ while is $0$ otherwise. 
In particular, for $i=k$, $\ket{\scriptstyle{2^i-1-\sum_{h=1}^{k-1}2^{h}}}=\ket{0\cdots01}$ and therefore $_a\mel{1}{U_{k-1}}{z}^{\otimes d}_a=z^k$ which means $B_{k-1}=z^k$.
The recursive procedure consists of proving that $\mel{\scriptstyle{2^i-1-\sum_{h=1}^{k}2^{h}}}{U_{k}}{z}^{\otimes d}_a=z^i$ starting from the assumption in the Equation \ref{eq:as2}. 
Let's start from the state $U_{k-1}\ket{z}^{\otimes d}_a$ and let's apply on it the transformation $C_{a_0}X_{a_k}\Bar{C}_{a_k}R_y(-2\vartheta_{k-1})_{a_0}$. 
The transformation $\Bar{C}_{a_k}R_y(-2\vartheta_{k-1})_{a_0}$ acts only on the state $\ket{s_{d-1}s_{d-2}\cdots s_0}$ where $s_{k}=0$, therefore, it does not act on the states $\ket{\scriptstyle{2^i-1-\sum_{h=1}^{k-1}2^{h}}}$ $\forall i>k$. 
Instead $C_{a_0}X_{a_k}$ is a bit-flip transformation which acts on the state $\ket{s_{d-1}s_{d-2}\cdots s_0}$ only if $s_{0}=1$ and it applies a NOT operation on the bit $s_k$, therefore 
\begin{equation}
    C_{a_0}X_{a_k}\ket{\scriptstyle{2^i-1-\sum_{h=1}^{k-1}2^{h}}}=\ket{\scriptstyle{2^i-1-\sum_{h=1}^{k}2^{h}}} 
\end{equation}
That means 
\begin{equation*}
    \mel{\scriptstyle{2^i-1-\sum_{h=1}^{k}2^{h}}}{U_{k}(\vec\vartheta_{k-1})}{z}=z^i
\end{equation*}
for $d>i>k$ and, in particular, for $i=k+1$, $_a\mel{1}{U_{k-1}}{z}^{\otimes d}_a=z^k$ therefore $B_{k}=z^{k+1}$.
Such final result proves that $B_{k}=z^{k+1}$ $\forall k=0,\dots,d-1$, therefore
\begin{equation*}
    _a\mel{0}{U_{k}}{z}^{\otimes d}_a=f_{k-1}(z)\cos{\vartheta_{k-1}}-B_{k-1}\sin{\vartheta_{k-1}}=f_k(z)
\end{equation*}
The second step is therefore concluded and thus the proof of the theorem.

\qquad\hspace*{\fill}$\square$ \\
In the Figure \ref{ref:fig4}\textbf{-a}, $S_U$ is the
subroutine which achieves the second step of the perceptron algorithm, the composition of
the polynomial expansion in $z$, and it is equal to $U_d(\vec\vartheta_{d-1})$.


\section{Approximation of analytical activation functions}
The transformation $S_U=U_d\otimes X^{\otimes n}$ applied on $|\psi^d_{z}\rangle$ returns a
state with the probability amplitude associated to $\ket{0}_a\ket{0}_q$ equal to
$2^{-d/2}f_d(z)$.  Let's denote such final quantum state of $(n+d)$-qubits as
$|\psi_{f(z)}^d\rangle$.  Eq. \ref{eq:ric2} defines $f_d(z)$ as a $d$-degree polynomial with
coefficients depending on $d$ angles $\vartheta_k,\,k=0,\ldots,d-1$. From Theorem
\ref{th:teof}, there follows a Corollary which shows how to set such angles in order to
approximate an arbitrary analytical activation function $f(z)$ by $f_d(z)$.

\begin{corollary}\label{co:co2}
  Let $f$ be a real analytic function over a compact interval $I$. If $f_d$ is the top
  member of the family of polynomials defined in Theorem \ref{th:teof} (Eq. \ref{eq:ric2})
  then the angles $\vartheta_k,\,k=0,\ldots,d-1$ can be chosen in such a way that $f_d(z)$
  coincides with the $d$-order Taylor expansion of $f(z)$ around $z=0$, up to a constant
  factor $C_d$ which depends on $f$ and on the order $d$ as 
\begin{equation}\label{eq:-cd}
        C_d=a_k\prod_{j=k}^{d-1}\left(\cos{\vartheta_j}\right)^{-1} \;,
\end{equation}
where $a_k=\frac{1}{k!}f^{(k)}(0)$ is the first non-zero coefficient of the expansion and $f^{(k)}$
the $k$-order derivative of $f$.
\end{corollary}
\textbf{Proof:} 
Let's denote with $ T_d(z)$ the truncated polynomial series expansion of an analytical function $f$ at the order $d$ expressed by $ T_d(z)=\sum_{i=0}^{d}a_i z^i$.\\
Let $f$ be an analytical function then it exists $k$, where $0\leq k\leq d$ and $a_i=0 \hspace{2mm}\forall i< k$, such that, factorizing $a_k$, $T_d(z)$ results 
   \begin{equation}
      T_d(z)= a_k\left[z^k+\sum^{d-1}_{i=k}\frac{a_{i+1}}{a_k}z^{i+1} \right]  
    \end{equation}
Let's consider the value $f_d(z)$, defined by the Equation \ref{eq:ric2}. If $ \cos{\vartheta_i}\neq 0$ $\forall i=k,\dots,d-1$ and $ \vartheta_i=-\frac{\pi}{2}$ otherwise, then, factorizing any $ \cos{\vartheta_i}\neq 0$, it can be expressed as 
    \begin{equation}
      f_d(z)=A_{dk}\left[z^k- \sum_{i=k}^{d-1}\frac{\tan{\vartheta_{i}}}{A_{ik}}z^{i+1}\right]
    \end{equation}
where $\scriptstyle A_{ik}=\prod_{j=k}^{i-1}\cos{\vartheta_j}$.    
Let's choose the angles $ \vartheta_i$ in order to satisfy the equality
\begin{equation} T_d(z)=\frac{a_k}{A_{dk}}f_d(z)
\label{eq:diffunc}
\end{equation}
Therefore, equalizing term by term in powers of $z$, the resulted equation for $i\geq k$ is 
\begin{equation}
     \vartheta_{i} = \arctan\left(-\frac{a_{i+1}}{a_k}A_{ik}\right)
    \label{eq:trovatheta}
\end{equation}
Since the values $A_{ik}$ depend by the angles $\vartheta_k,\dots,\vartheta_{i-1}$ then the angles $\vartheta_i$ in turn depend on them. It means that the computation of all the angles must been ordered from $\vartheta_k$ to $\vartheta_{d-1}$.
From such definition of the angles $\vartheta_i$, where $i=0,\dots,d-1$, the Equation \ref{eq:diffunc} is satisfied. Therefore $f_d(z)$ is equal to the series expansion of $f(z)$ at the order $d$ less than a constant factor $C_d$
\begin{equation}
    C_d=\frac{a_k}{A_{dk}}=\frac{a_k}{\prod_{j=k}^{d-1}\cos{\vartheta_j}}
\end{equation}
Its value is constant while $z$ changes, and it depends on the coefficients $a_i$ of the Taylor expansion of the function $f$ where $i=k,\dots ,d$.
\\\qquad\hspace*{\fill}$\square$\\


\section{Computation of the amplitude}
To summarize, the quantum circuit so far defined employs $n+d$ qubits and it performs two
transformations: the first sends the state $\ket{0}_a^{\otimes d}\ket{0}_q^{\otimes n}$
into $|\psi_z^d\rangle$, as a consequence of Theorem \ref{th:3teo}, while the second is
$ S_U\otimes X^{\otimes n}$ which returns a state having $2^{-d/2}f_d(z)$ as probability
amplitude corresponding to the state $\ket{0}_a\ket{0}_q$.  An
important property of such quantum circuit, which is shown in the Figure
\ref{ref:fig4}-\textbf{a}, is that it encodes the value $f_d(z)$ (up to the constant
$2^{-d/2}$), which is non-linear with respects to the input values $\vec x$, in a quantum
state (the state $|\psi^d_{f(z)}\rangle$).  Indeed, in a generic context, the
quantum circuit in Figure \ref{ref:fig4}-\textbf{a} can be integrated in a circuit for a
multi-layer qubit-based neural network.  In such context the value $f_d(z)$ corresponds to
the output value of a hidden neuron.  The freedom left by the non-destroying activation
function makes possible to build a deep qubit-based neural network. Each new layer
receives quantum states, like the state prepared to enter the network at the first layer.
As last result, here we explicitly show how to operate the last layer of the network, by focusing
on the case of a one-layer perceptron. 

While the circuit in Figure
\ref{ref:fig4}-\textbf{a} returns a state $|\psi^d_{f(z)}\rangle$ which has the
value $2^{-d/2}f_d(z)$ encoded as a probability amplitude, the circuit in
Figure \ref{ref:fig4}-\textbf{b} allows to estimate such amplitude.  It implements a qubit-based version of a
one-layer perceptron.  
Any quantum algorithm ends by extracting information from the quantum state of the qubits
by measurement operations.  Applying the measurement operations to the qubits of the circuit of Figure
\ref{ref:fig4}-\textbf{a} allows to estimate only the probability to measure a given
quantum state, but from the probability it is not possible to compute the inherent
amplitude. Indeed, the probability is the square module of the amplitude, therefore, it
does not preserve the information about the phase factor (the sign for real values) of the
amplitude.
To achieve such a goal a straightforward method consists in defining a quantum circuit which returns a quantum state with a value $\frac{1}{2}(1+2^{-d/2}f_d(z))$ stored as a probability amplitude, a task operated by the circuit in the Figure \ref{ref:fig4}-\textbf{b}.\\
Such quantum circuit operates on a register $l$ of a single qubit, in addition to the
registers $q$ and $a$, and it returns the probability $\frac{1}{4}|1+2^{-d/2}f_d(z)|^2$ to observe the state $\ket{0}_l\ket{0}_a\ket{0}_q$.  
Let's now focus on such circuit devoted to the estimation of the perceptron output.
Let's consider an $n$-qubits state $\ket{\psi}$ with real amplitudes and an unitary operator $U$ such that $ U\ket{0}^{\otimes n}=\ket{\psi}$. In order to estimate the amplitude $\braket{0}{\psi}$, where  $\ket{
0} \equiv \ket{0}^{\otimes n}$, a three-step algorithm can be defined to achieve the goal.
The algorithm foresees the use of $(n+1)$-qubits, $n$ of which to store $\ket{\psi}$ labeled with $q$ and one additional qubit $l$.
The said three steps consist of a Hadamard gate on $l$, the transformation $U$ applied to the qubits of the register $q$ and controlled by $l$ and another Hadamard gate on $l$.
Indeed, starting from the state $\ket{0}_l\ket{0}_q^{\otimes n}$, after the first Hadamard gate the $(n+1)$-qubits state becomes $\frac{1}{\sqrt{2}}\left(\ket{0}+\ket{1}\right)\ket{0}^{\otimes n}$.\\
With the controlled-$U$ transformation the state becomes $\frac{1}{\sqrt{2}}\left(\ket{0}\ket{0}^{\otimes n}+\ket{1}\ket{\psi}\right)$ and, with the last Hadamard gate, $\frac{1}{2}\left[\ket{0}\left(\ket{0}^{\otimes n}+\ket{\psi}\right)+\ket{1}\left(\ket{0}^{\otimes n}-\ket{\psi}\right)\right]$. 
After a measurement of the $n+1$ qubits, the probability to measure the state $\ket{0}_l\ket{0}_q^{\otimes n}$ is $ P_{0}=\frac{1}{4}\abs{1+\braket{0}{\psi}}^2$.
After the estimation of $P_0$ the amplitude $\braket{0}{\psi}$ is achievable by reversing the formula, therefore $\braket{0}{\psi}=2\sqrt{P_0}-1$. 
The square module is invertible in such case because $\abs{\braket{0}{\psi}}\leq 1$.
Let's apply such a method of amplitude estimation in the case of the quantum perceptron.
The quantum circuit, exposed in the previous sections and shown in the Figure \ref{ref:fig4}-\textbf{a}, applied on two qubit registers $q$ and $a$, each initialized in the state $\ket{0}$, it returns an $(n+d)$-qubits state $\scriptstyle\ket{\psi^d_{f(z)}}$. 
The circuit is summarized as a series of $X$ gates applied on the qubits in the register $q$, the subroutine $S_V$ applied on the two qubit registers followed by $S_U$ applied on the register $a$ and another series of $X$ gates applied on the qubits in the register $q$.
The circuit turns out $ X^{\otimes n}S_US_V X^{\otimes n}\ket{0}^{\otimes d}\ket{0}^{\otimes n}=\scriptstyle\ket{\psi^d_{f(z)}}$.
Therefore, to estimate
${_q}\!\bra{0}{_a}\!\bra{0}\scriptstyle\ket{\psi^d_{f(z)}}=2^{-d/2}f_d(z)$, let's
apply the amplitude estimation algorithm described above where the transformation $U$ is $
X^{\otimes n}S_US_V X^{\otimes n}$ and, in turn, the state $\ket{\psi}$ is the state
$\ket\psi^d_{f(z)}$.  Therefore, after a measurement all over the qubits, the probability
to obtain the state $\ket{0}_l\ket{0}_a^{\otimes d}\ket{0}_q^{\otimes n}$ is
\begin{equation}
    P_0=\frac{1}{4}\left|1+2^{-d/2}f_d(z)\right|^2
    \label{eq:firstll}
\end{equation}
From the estimation of $P_0$ it is possible to compute an estimation of $2^{-d/2}f_d(z)$.
Summarizing, the circuit, which allows to estimate the amplitude $2^{-d/2}f_d(z)$, is $H_lC_l(X^{\otimes n}S_US_VX^{\otimes n})H_l$
with a measurement for each qubits in the registers $q$, $a$ and $l$, respectively. Notice that such quantum circuit is partially different with respect from the circuit in the Figure \ref{ref:fig4}\textbf{-b}.
As from Figure \ref{ref:fig4}\textbf{-b} the subroutine $S_V$ is not controlled by $l$. Indeed $S_V$ is built such that $ S_V\ket{0}^{\otimes d}\ket{0}^{\otimes n}=\ket{0}^{\otimes d}\ket{0}^{\otimes n}$. Therefore, the circuit $H_lC_l(X^{\otimes n}S_US_VX^{\otimes n})H_l$ and $H_lC_l(X^{\otimes n}S_U)S_V C_l(X^{\otimes n})H_l$ allows to achieve the same purpose of building a state with $P_0$ as probability to obtain the state $\ket{0}_l\ket{0}_a^{\otimes d}\ket{0}_q^{\otimes n}$ after a measurement for each qubits.

Let's remark that Theorems \ref{th:3teo} and \ref{th:teof} imply that a quantum state with
$2^{-d/2}f_d(z)$ as superposition coefficient does exist and, since any quantum state
$\ket{\psi}$ is normalized then $2^{-d/2}f_d(z)\leq 1$.
From such results, by defining $P_0=\frac{1}{4}|1+2^{-d/2}f_d(z)|^2$ it is possible to reverse the equation to find $f_d(z)$, once given the probability $P_0$.

Therefore, the quantum perceptron algorithm consists of an estimation of the probability
$P_0$ feasible with a number $s$ of measurement operations of all the qubits. The error
over the estimation of $P_0$ depends on the number of samples $s$.  The resulting output
of the qubit-based perceptron  is written
\begin{equation}\label{eq:outputy}
  y_q=2^{d/2}(2\sqrt{P}-1)C_d
\end{equation}
where $P$ is the estimation of $P_0$ and $C_d$ is defined in the Equation \ref{eq:-cd}.
Hence $y_q$ provide the estimation of the value $f_d(z)$, which is the
polynomial expansion of the activation function $f$ at the order $d$.  Once the estimation
of $P_0$ is obtained by a quantum computation, the value $y_q$ is derived by a classical
computation. The estimation of $P_0$ is given by $P=m/S$ where $S$ is the total number of the
measurements of $\ketbra{0}_l\otimes\ketbra{0}_a\otimes\ketbra{0}_q$ and $m$ is the number of those measurements  
which return $1$ as result. 
A second way to estimate $y_q$ is given by the quantum amplitude estimation algorithm \cite{brassard2002quantum}.
Briefly, let's consider a transformation $\mathcal{A}$ such that $\mathcal{A}\ket{0}=\ket{\Psi}=\sqrt{1-a}\ket{\psi_0}+\sqrt{a}\ket{\psi_1}$ where $\ket{\psi_0}$ and $\ket{\psi_1}$ are $n$-qubits states and $a\in\left[0,1\right]$, the quantum amplitude estimation algorithm computes, with an additional register of $m$ qubits, a value $\Tilde{a}$ such that at most $\abs{\Tilde{a}-a}\sim\mathcal{O}(M^{-1})$ where $M=2^m$. Therefore, to apply to the case under consideration namely $X^{\otimes n}S_US_V X^{\otimes n}\ket{0}^{\otimes d}\ket{0}^{\otimes n}=|\psi_{f(z)}^d\rangle$, one may take  $\mathcal{A}=X^{\otimes n}S_US_V X^{\otimes n}$, $\ket{\Psi}=|\psi_{f(z)}^d\rangle$ being the latter a state over $n+d$ qubits, and $\sqrt{a}=2^{-d/2}f_d(z)$, respectively, so that the output of the perceptron results as
\begin{equation}
    y_q=2^{d/2}\sqrt{\Tilde{a}}C_d-\gamma
    \label{eq:qaea}
\end{equation}
where $\gamma$ is the shift value used to make the function $f_d(z)\in\left[0,1\right]$ within $z\in\left[-1,1\right]$.

\section{Discussion}

The results derived above allow to implement a multilayered perceptron of arbitrary size in
terms of neurons per layer as well as number of layers. To test it, one may restrict the
quantum algorithm to a one-layer perceptron. The tensorial calculus to obtain both $z$ and
$f_d(z)$ has to be performed by a quantum computer and eventually ends with a probability
estimation.  The estimation $P$ of the probability, accordingly over $y_q$, is subject to two
distinct kinds of errors: one depending on the quantum hardware and a random error due to
the statistics of $m$ which is hardware--independent.  

In the following discussions, the
error is analyzed as a function of the required qubits ($n$ and $d$) and the number of
samples $S$. The case of a qubit-based one-layer perceptron with $N_{in}=4$ is then
explored in order to verify the capability of the algorithm to approximate $f(z)$.
Because of the significant number of quantum gates involved, a quantum simulator has been
used.

\subsection{Error estimation and analysis}

From Equation \ref{eq:outputy}, $y_q$ depends on $P$, the estimation of the probability
that a measurement of $\ketbra{0}_l\otimes\ketbra{0}_a\otimes\ketbra{0}_q$ returns $1$ as
a result. The estimation $P$ is subject to a random error.
Indeed $P=m/S$, where the number of successes $m$ is a random variable that follows the
binomial distribution 
\begin{equation*}
  B(m,S,P_0) = {S \choose m} P_0^m(1-P_0)^{S-m}\;.
\end{equation*}
Therefore $Var[m]=SP_0(1-P_0) \le\tfrac14 S$ and $Var[P]=P_0(1-P_0)/S
\le\tfrac1{4S}$. Then Eq.~\ref{eq:outputy} implies that
\begin{equation}\label{eq:varsigy}
    \sigma_{y_q}\propto\sqrt{\frac{2^d}{S}}C_d
\end{equation}
where $\sigma_{y_q}=(Var[y_q])^{1/2}$.  This means that $s$ must grow at least as $2^d$
in order to hold $\sigma_{y_q}$ constant in $d$. Since also $C_d$ increases with $d$, $S$
should grow even faster.  The error
$\sigma_{y_q}$ is clearly hardware independent.  The implementation on a quantum device,
rather than a simulator, requests further analysis of the hardware-dependent errors. 

Multi-qubit operations on a quantum device are subject to two kind of errors \cite{cross2019validating}\cite{tannu2019not}: namely the
limited coherence times \cite{bouchiat1998quantum} of the qubits and the physical
implementations of the gates \cite{girvin2011circuit}.
Therefore the number of gates (the circuit depth) has a double influence on the error: a large number of gates requires a longer circuit run time which is limited by the coherence times of the qubits moreover each gate introduces an error due to its fidelity with respect to the ideal gate.
The evaluation of the number of gates gives an estimate of the hardware-dependent error.
The number of gates of the proposed qubit--based perceptron algorithm is
$\sim 330$ for $d=1$ and it increases by $\sim 400$ when $d$ increases by $1$, in other words almost linearly with $d$.
Instead, the number of gates depends exponentially on  $n$ \cite{shende2006synthesis}. Depending on the size of the problem and the maturity of the technology to implement the physical qubits, a real application should  require to use logical qubits or at least to integrate quantum error correction coding embedded within the algorithm.
These prerequisites are especially necessary if the quantum amplitude estimation algorithm is applied to estimate $y_q$ as in Eq. \ref{eq:qaea}.
In such algorithm the depth of the circuit increases as $\mathcal{O}(MNd)$ instead $\mathcal{O}(Nd)$ of the vanilla method of the Eq.\ref{eq:outputy}, but the error over the estimation of $y_q$ decreases with $M^{-1}$. Indeed, since at most $\abs{\Tilde{a}-a}\sim\mathcal{O}(M^{-1})$ and the error over the average of $\Tilde{a}$ computed with $S$ shots, is at most $\frac{1}{\sqrt{S}M}$, from the error propagation of the Eq. \ref{eq:qaea} one has
\begin{equation}
    \sigma_{y_q}\propto\sqrt{\frac{2^d}{S}}\frac{C_d}{M}
\end{equation}
where $M=2^m$. Therefore the error can be reduced exponentially by increasing $m$, the number of additional qubits of the quantum amplitude estimation algorithm.
A further analysis of the impact of the noise of a quantum hardware on the output of the perceptron is present in the Supplementary Note \textbf{3 }, where the results of the calculations obtained with a simulated noise model are presented.



\begin{figure*}[t!]
  \centering
  \includegraphics[scale=.75]{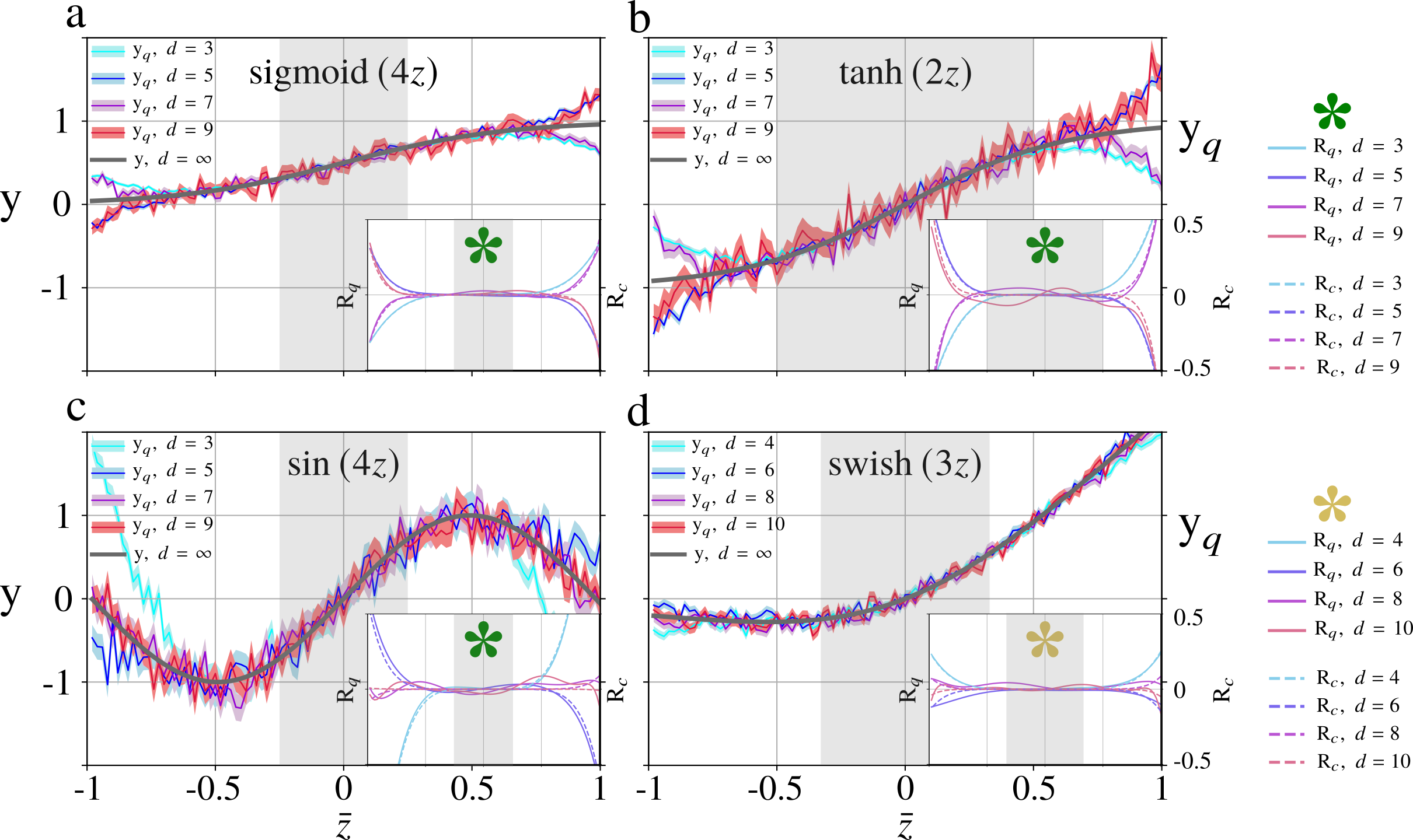}
  \caption{\textbf{Enlarged view of the output of a quantum one-layer perceptron with different analytical activation functions.} The output of the classical perceptron $y$ (gray lines) as a function of $\bar z$  is computed with four different target activation functions $f$: a hyperbolic tangent $tanh(z)$ \textbf{(a)}, a logistic function $sigmoid(z)$ \textbf{(b)}, a sine function $\sin(z)$ \textbf{(c)} and the swish function $z\cdot sigmoid(z)$ \textbf{(d)}. 
		The estimation of the quantum perceptron output $y_q$ approximates $y$ at the polynomial expansion's order of $d=3$ (cyan line) computed with $S=2^{16}$ samples, $d=5$ (blue line) with $S=2^{18}$, $d=7$ (violet line) with $S=2^{20}$  and $d=9$ (red line) with $S=2^{22}$ in the cases \textbf{(a)}, \textbf{(b)} and \textbf{(c)} while for the swish function the same colors are of one extra order ($d+1$) and the same number of samples. In the lower right graphs the plots of $R_c$ and $R_q$ are shown in function of $\bar z$ and at different order $d$. The gray regions delimit the interval of $\bar z$ rescaled by the factor $k$ which is the relevant region of interest to behave as monotone activation function. 
		}
	\label{fig3}
\end{figure*}

\subsection{Implementation of a quantum one-layer perceptron with $N_{in}=4$}

In order to show a practical example, we consider now the implementation
of a quantum one-layer perceptron with $4$ input neurons. With $N_{in}=4$, $z$ is computed
with $n=3$ qubits.  To test the algorithm, the analytic activation functions $f$ to be
approximated are the hyperbolic tangent, the sigmoid, the sine and the swish function
\cite{ramachandran2017searching} respectively (Fig. \ref{Fig3}).  To estimate the
perceptron output $y_q$, $n+d+1$ qubits are required where $d$ is the order of
approximation of the activation function $f$.
The output $y_q$ is computed following Eq.~\ref{eq:outputy} where the probability $P$ is
estimated with the measurement of $s$ copies of the quantum circuit in the Figure \ref{ref:fig4}-\textbf{b}.  To evaluate the effectiveness of the algorithm in reconstructing
the chosen activation function $f$, the output $y_q$ is compared with
$f(z)$ at different values of the inputs $x_i$, the weights $w_i$ and the bias $b$, where
$i=0,1,2,3$.
We extend the evaluation on different activation functions also to 
different orders $d$, in order to exhaustively check the algorithm.

The weight vector is set to $\vec w=(1,1,1,1)$ and the bias to $b=0$. The input vector $\vec x$ varies as $\vec x = \bar z(1,1,1,1)$ with $\bar z\in[-1,1]$.
In order to make the approximation capability manifest, since plotting as a function of $\bar z$ the activation functions does not differ significantly from their linear approximation, we consider a rescaled horizontal axis by  $f(kz)$ with $k=4$ for the sigmoid and the sine function, $k=3$ for the swish function and $k=2$ for the hyperbolic tangent. In Figure \ref{fig3} the different activation functions $f$ are plotted versus $\bar z = \frac{5}{4}z\in[-1,1]$. 
The rescaled factor here employed only to better show the effectiveness of the algorithm, can be set arbitrarily through the computation  of the angles $\vartheta$, according to Theorem \ref{th:teof}, by simply
considering the Taylor expansions of $f(kz)$. 
The case of random weight vectors and biases are reported in the Supplementary Figures.  

The quantum perceptron algorithm has been developed in Python using the open--source quantum computing
framework QisKit.
The quantum circuit was run on the local quantum simulator \textit{qasm\_simulator}
available in the QisKit framework.
Using a simulator, the only error over the estimated value $y_q$ is $\sigma_{y_q}$. To keep $\sigma_{y_q}$ constant, the
number of samples $S$ must increase with $d$ to compensate the factor
$2^{d/2}C_d$ in Eq.~\ref{eq:varsigy}. Figure \ref{fig3} was obtained with the starting 
choice $S=2^{16}$ and $d=3$.

To evaluate how well $y_q$ approximates the $d-$ order Taylor expansion of a given activation function, let's define $R_q=y-\Tilde{y}_q$, where $y=f(z)$ and $\Tilde{y}_q$ is its polynomial fit. The values $R_q$ are compared with $R_c=y-T_d$, where $T_d$ is the Taylor expansion of order $d$.
For $k=1$, the full Taylor series of the activation functions under study converge in the interval $\left[-1,1\right]$ of $\bar z$. However, only in the case of the sine function the convergence holds true all over $\mathbb{R}$. As a consequence, for $f(z) = \sin(4z)$, $R_c$ goes to $0$ for any value of $\bar z$ when $d$ increases. For the other functions in Figure \ref{fig3} the convergence radius of the polynomial series is finite and it depends on $k$. In the case of $\tanh(2z)$, for instance, the convergence radius is less than $1$ and $R_c$ does not decrease with $d$ for $\bar z$ large enough. Therefore, it is more representative to compare $R_q$ with $R_c$ rather than the activation function itself.

To quantify the difference between $R_q$ and $R_c$, let's compute the mean square error (MSE) at different orders $d$ in the gray region of $\bar z$ in the Figure \ref{fig3}.
The value of MSE between $R_q$ and $R_c$ is almost always of the order $10^{-5}$ for $d=3$ ($4$ for the swish) and $10^{-4}$ when $d$ increases. For the hyperbolic tangent the MSE is of the order $10^{-3}$ when $d=7$ and $9$.

The reason of such difference between $\tanh$ and the other functions is due to the trend of $C_d$ with $d$.
Indeed $C_d$ increases with
$d$, therefore, it is not sufficient to duplicate $S$ when $d$ increases by $1$ because $C_d$
becomes relevant. Such an effect is more evident if the rescale factor $k$ is high.
Indeed, for the examined functions, with $k=1$ the contribution of $C_d$ to the error of $y_q$ is not relevant also for a high value of $d$, because $C_d$ increase as a polynomial with $d$ if $k=1$. 
At higher values of $k$, $C_d$ increases exponentially with $d$ and it gives a relevant contribution to the error.   

	\section{Conclusions}
A $n-to-2^n$ quantum perceptron approach is developed  with the aim of implementing a general and flexible quantum activation function, capable to reproduce any standard classical activation function on a circuital quantum computer.  Such approach leads to define a truly quantum Rosenblatt perceptron, scalable to multi-layered quantum perceptrons, by having prevented the need of performing a measurement to implement non-linearities in the algorithm. 
To conclude, our quantum perceptron algorithm fills the lack of a method to create arbitrary activation functions on a quantum computer, by approximating any analytic activation functions to any given order of its power series, with continuous values as input, weights and biases. 
Unlike previous proposals, we have shown how to approximate any analytic function at arbitrary approximation, without the need to measure the states encoding the information. 
By construction, the algorithm bridges quantum neural networks implemented on quantum computers with the requirements to enable universal approximation as from Hornik's theorem. Our results pave the way towards mathematically grounded quantum machine learning based on quantum neural networks.

	
	\section{Data availability}
	The data that support the findings of this study are available from the corresponding author upon reasonable request.
	
	\section{Competing interest}
	The authors declare that there are no competing interests
	

	\newpage
	
	\bibliography{\jobname} 
	\bibliographystyle{naturemag}

\end{document}